\documentclass[12pt]{iopart}
\usepackage{iopams,setstack}
\usepackage{subeqnarray}
\usepackage[dvips]{graphicx,color}

\newcommand{\ket}[1]{|#1\rangle}         
\newcommand{\bra}[1]{\langle#1|}         
\newcommand{\cats}[1]{\ensuremath{\big|#1\big>}}
\newcommand{\bras}[1]{\ensuremath{\big<#1\big|}}

\newcommand{\id}{\mathrm{id}}           

\newcommand{\BEQ}{\begin{equation}}     
\newcommand{\BEA}{\begin{eqnarray}}
\newcommand{\EEQ}{\end{equation}}       
\newcommand{\EEA}{\end{eqnarray}}
\newcommand{\eps}{\varepsilon}          

\renewcommand{\vec}[1]{\boldsymbol{#1}} 

\newenvironment{smallmatrix}{\left(\begin{array}{cc}}{\end{array}\right)}
\newcommand{\dotsc}{\ldots}

\newcommand{\fig}[3]{\begin{figure}[!ht]%
    \centering\includegraphics[width=0.6\textwidth]{#1}%
    \caption{#3}\label{#2}\end{figure}}

\begin{document}

\title[Ageing without detailed balance]{Ageing phenomena without 
detailed balance: the contact process}
\author{Tilman Enss$^a$, Malte Henkel$^b$, Alan Picone$^b$ 
and Ulrich Schollw\"ock$^c$}
\address{$^a$Max-Planck-Institut f\"ur Festk\"orperforschung, 
Heisenbergstr.\ 1, D--70569 Stuttgart, Germany}
\address{$^b$Laboratoire de Physique des 
Mat\'eriaux\footnote{Laboratoire associ\'e
au CNRS UMR 7556}, Universit\'e Henri Poincar\'e Nancy I, B.P.\ 239,
F--54506 Vand{\oe}uvre-l\`es-Nancy Cedex, France}
\address{$^c$Institut f\"ur Theoretische Physik C, RWTH Aachen, 
D--52056 Aachen, Germany}
\eads{\mailto{Tilman.Enss@fkf.mpg.de}, 
\mailto{henkel@lpm.u-nancy.fr}, 
\mailto{picone@lpm.u-nancy.fr}, 
\mailto{scholl@physik.rwth-aachen.de}}
\begin{abstract}
The long-time dynamics of the $1D$ contact process suddenly brought out of 
an uncorrelated initial state is studied through a light-cone transfer-matrix
renormalisation group approach. At criticality, the system 
undergoes ageing which is characterised through the dynamical scaling of the
two-times autocorrelation and autoresponse functions. The observed non-equality
of the ageing exponents $a$ and $b$ excludes the possibility of a finite
fluctuation-dissipation ratio in the ageing regime. The scaling form of the
critical autoresponse function is in agreement with the prediction of 
local scale-invariance. 
\end{abstract}
\submitto{\JPA}
\date{3$^{\rm rd}$ of June 2004}

\section{Introduction}
\label{sec:intro}

Consider a statistical system prepared in some initial state. How does
it relax towards one of its stationary states? This problem was first
studied systematically in glassy systems where is was observed that
the approach towards the thermodynamic equilibrium in this kind of
system can be very slow (formally the relaxation time becomes infinitely
large) and depends on the details of the history how the relaxing
material was treated. While this seemed to preclude any systematic
study of such systems, it was found empirically \cite{Stru78} that the
time-dependence of observables can be cast into dynamical scaling
forms such that universal and reproducible properties of the
relaxation process emerge. Later, similar effects were also observed
to occur in spin and structural glasses (see e.g. \cite{Bouc00}) 
and in non-glassy, e.g.\ simple ferromagnetic,
systems.\footnote{In these studies, it is always assumed that the
underlying dynamics satisfies detailed balance.} From a microscopic
point of view, the relaxation in these systems proceeds through the
formation of correlated domains of a time-dependent typical size
$\ell(t)$. This slow motion arises because any local discrete spin
variable will have at least two distinct equilibrium values it could
relax into. For a given position $\vec{r}$, the locally fluctuating 
variables (e.g.\ a local magnetic field $h(\vec{r})$ in magnetic
systems) will select towards which of the possible equilibrium states
the local order parameter $\phi(\vec{r})$ will evolve but there does
remain a competition between the distinct equilibrium states at the
macroscopic level. For simple ferromagnets and other non-glassy
systems, it is generally admitted that $\ell(t)\sim t^{1/z}$, where
$z$ is the dynamical exponent. For systems relaxing towards an {\em
equilibrium} steady-state and which are brought infinitely rapidly
from their initial state into contact with a thermal bath of temperature
$T$, the value of $z$ depends on whether $T<T_c$ or $T=T_c$, where
$T_c$ is the equilibrium critical temperature of the system. It is
convenient to study these relaxation phenomena through two-time
quantities such as the two-time autocorrelator $C(t,s)$ and the two-time
linear autoresponse function $R(t,s)$ defined by
\BEQ
C(t,s) = \left\langle \phi(t) \phi(s)\right\rangle \;\; , \;\;
R(t,s) = \left.\frac{\delta \langle\phi(t)\rangle}{\delta h(s)}\right|_{h=0}
\EEQ
where $\phi(t)$ is the time-dependent order parameter 
and $h(s)$ is the magnetic
field conjugate to $\phi$. Causality implies that $R(t,s)=0$ for $t<s$.
By definition, a system is said to undergo {\em ageing}, if either $C(t,s)$
or $R(t,s)$ does not merely depend on the time difference $\tau=t-s$, but 
on both the {\em observation time} $t$ and the {\em waiting time} $s$. 
Ageing occurs for quenches to temperatures either below or at the critical 
temperature but systems in the high-temperature phase with $T>T_c$ do not age. 
For recent reviews, see e.g.\ \cite{Bouc00,Bray94,Cugl02,Godr02,Cris03,Henk04}. 

Ageing systems may display dynamical scaling in the long-time limit
\cite{Stru78,Bouc00,Bray94,Cugl02,Godr02,Cris03,Henk04}. Specifically,
consider the two-time functions in the ageing regime $t\gg t_{\rm micro}$,
$s\gg t_{\rm micro}$ and $\tau =t-s\gg t_{\rm micro}$, where $t_{\rm micro}$
is some microscopic time. Then one expects the scaling behaviour
\BEQ \label{1:gl:CRskal}
C(t,s) \sim s^{-b} f_{C}(t/s) \;\; , \;\;
R(t,s) \sim s^{-1-a} f_{R}(t/s)
\EEQ
where the scaling functions $f_{C,R}(x)$ have the following asymptotic
behaviour for $x\to\infty$
\BEQ \label{1:gl:fCR}
f_{C}(x) \sim x^{-\lambda_C/z} \;\; , \;\;
f_{R}(x) \sim x^{-\lambda_R/z}
\EEQ
Here $\lambda_C$ and $\lambda_R$ are called the autocorrelation 
\cite{Fish88,Huse89} and autoresponse \cite{Pico02} exponents, respectively. 
The values of the exponents
$\lambda_{C,R}$ and $z$ depend on whether $T<T_c$ or $T=T_c$.
For example, $z=2$ for $T<T_c$ and a non-conserved order parameter.
In general, the exponents $\lambda_C$ and $\lambda_R$ are distinct, but for an
infinite-temperature initial state it can be shown that Galilei-invariance
of the model at zero temperature is a sufficient and model-independent 
criterion for the equality $\lambda_C=\lambda_R$ \cite{Pico04}. Indeed,
for a disordered initial state, $\lambda_C=\lambda_R$ had been taken for 
granted since a long time, see e.g. \cite{Bouc00,Bray94,Godr02}, 
and for recent reconfirmations in interacting field-theory see
\cite{Cala03,Maze04}. In that case, one
has $\lambda_C\geq d/2$ \cite{Yeun96}. The exponents $\lambda_{C,R}$ are 
independent of the equilibrium exponents and of $z$ \cite{Jans89}.

For ageing ferromagnetic systems with a non-conserved order parameter,
the value of the exponent $a$ depends on the properties of the
equilibrium system as follows \cite{Henk02a,Henk03e}. A system is said
to be in {\em class S} if its order-parameter correlator 
$C_{\rm eq}(\vec{r})\sim \exp(-|\vec{r}|/\xi)$ with a finite $\xi$
and it is said to be in {\em class L}\footnote{For example, the kinetic
spherical model quenched to $T<T_c$ is in class L.} if 
$C_{\rm eq}(\vec{r})\sim |\vec{r}|^{-(d-2+\eta)}$, 
where $\eta$ is a standard equilibrium
critical exponent. Then
\BEQ
a = \left\{ \begin{array}{ll}
1/z          & \mbox{\rm ~~;~ for class S} \\
(d-2+\eta)/z & \mbox{\rm ~~;~ for class L}
\end{array} \right.
\EEQ
Furthermore, $b=0$ for $T<T_c$ and $b=a$ if $T=T_c$, 
see e.g.\ \cite{Godr02,Cris03,Henk04}. 
Systems quenched to $T=T_c$ are always in class L. 

The distance from equilibrium is conveniently measured through the
{\em fluctuation-dissipation ratio} \cite{Cugl94a,Cugl94b}
\BEQ \label{1:gl:FDR}
X(t,s) := T R(t,s) \left( \frac{\partial C(t,s)}{\partial s}\right)^{-1}
\EEQ
At equilibrium, the fluctuation-dissipation theorem states that $X(t,s)=1$. 
Ageing systems may also be characterised through the limit
fluctuation-dissipation ratio \cite{Godr00a,Godr00b,Godr02} 
\BEQ
\label{eq:xinf}
X_{\infty} = \lim_{s\to\infty} \left( \lim_{t\to\infty} X(t,s) \right)
\EEQ
Below criticality, one expects $X_{\infty}=0$, but if $T=T_c$, it should be
a universal number, and this has been confirmed in a large variety of systems
in one and two space 
dimensions \cite{Godr00b,Cala02a,Cala02b,Cala03,Henk03d,Sast03,Chat04}.
Alternatively, at $T=T_c$ one may fix $x=t/s$ and consider 
$X(x)=\lim_{s\to\infty}X(xs,s)$ and then $X_{\infty}=\lim_{x\to\infty} X(x)$.
The order of the limits is important, since 
$\lim_{t\to\infty}\left(\lim_{s\to\infty} X(t,s)\right)=1$.

The above discussion of the properties of ageing systems has implicitly
assumed that the system's dynamics satisfies detailed balance and therefore 
always relaxes towards an {\em equilibrium}
steady-state even if it may never reach it. Here we wish to
investigate the relaxation of more general systems where detailed
balance is no longer satisfied and whose steady-state therefore cannot
be in thermodynamic equilibrium. Probably the simplest kinetic system
fundamentally far from equilibrium is the celebrated {\em contact
process} which has a steady-state transition in the directed
percolation universality class.\footnote{See \cite{Mobi04} for an exactly
solvable kinetic Ising model with a non-equilibrium steady state.} 

In section~\ref{sec:cp}, we shall recall the definition of the model
and discuss the computation of correlators and response functions
before we present in section~\ref{sec:tmrg} those elements of the
light-cone transfer-matrix renormalisation group (LCTRMG) which are
important for our purposes.  Calculating two-time observables from the
LCTMRG, we study in section~\ref{sec:results} their time-dependent
scaling behaviour.  In particular, we shall investigate whether there
exist scaling forms analogously to
eqs.~(\ref{1:gl:CRskal},\ref{1:gl:fCR}) and if so, what the values of
the ageing exponents $a,b,\lambda_C/z,\lambda_R/z$ are.  We shall
inquire whether any analogue of a finite fluctuation-dissipation ratio
might exist. This question is of particular importance since in a
two-dimensional voter model an effective non-equilibrium temperature
can indeed be defined through eq.~(\ref{1:gl:FDR}) \cite{Sast03}.
However, the voter model still satisfies detailed balance (albeit in a
slightly unusual form) and we shall try and see whether the procedure
proposed in \cite{Sast03} might conceivably be extended to the contact
process. A more quantitative question concerns the form of the
time-dependent scaling functions. Indeed, for ageing ferromagnets it
has been shown that dynamical scaling as defined above can be extended
to a richer {\em local scale-invariance} \cite{Henk02}.  In
particular, the scaling forms of the two-time reponse function
$f_R(y)$ \cite{Henk01,Henk03b} and more recently also of the two-time
autocorrelation function $f_C(y)$ for $T<T_c$ \cite{Pico04,Henk04b}
have been derived.  We shall study whether an analogous generalisation
is possible in the contact process. A complementary paper studies the
same model through intensive Monte Carlo simulations \cite{Rama04}. 
In section~\ref{sec:conclusions}, we conclude.
    
\section{The contact process}
\label{sec:cp}

The $1D$ contact process (CP) is defined as follows and might be conceived as
a simple model describing the propagation of an epidemic disease, see 
\cite{Hinr00,Odor04} for reviews. Consider a chain where
each site can be either be empty (`healthy') or else be occupied by a particle 
of a single species $A$ (`infected'). The evolution occurs according to
the following rules and rates:
\BEA
  A\emptyset,\emptyset A 
       & \overset{\lambda}{\longrightarrow} & AA \mbox{\rm ~~(`infection')}  \\
  A    & \overset{1}{\longrightarrow}       & 0  \mbox{\rm ~~~~~(`healing')}.
\EEA

\fig{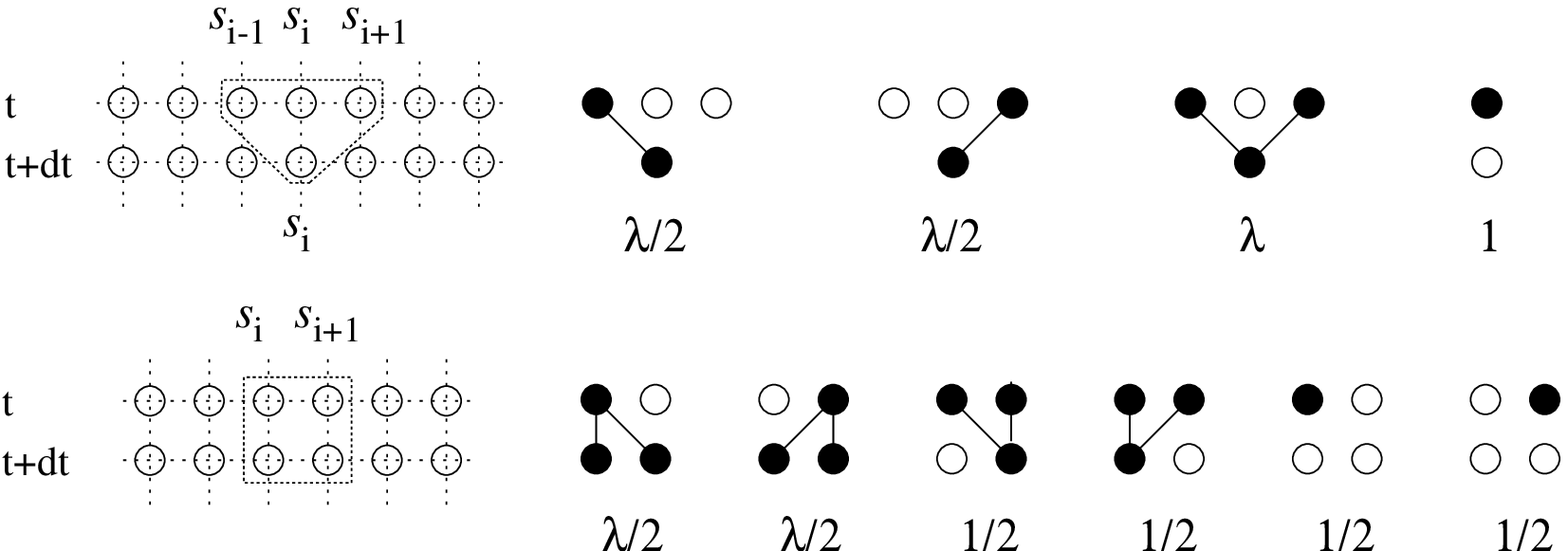}{fig:cpupdate}{Rules for the contact process on a $1D$ 
  lattice. Upper line: updating rules for a Monte Carlo simulation. Lower line:
updating rules for the LCTMRG.}

We shall consider throughout the case of one spatial dimension.  Then
an empty site next to an occupied one is infected with rate $\lambda$,
while infected sites heal at rate $1$, independently of their
neighbours.  In the upper panel of figure~1 we recall the elementary
processes for updating a $1D$ lattice with the corresponding rates and
this readily defines a Monte Carlo algorithm with asynchronous random
sequential updates. In the lower panel, we present a version of the CP
which is more adapted for use with the LCTMRG, see
section~\ref{sec:tmrg}.  Besides $\lambda$, it is also common to
parametrise the contact process with $p:=(1+\lambda)^{-1}$ as we shall
do in the following: if a lattice site $i$ is occupied, the particle
is annihilated with probability $p$, while if the site is empty, a
particle is created at one of the neighbouring sites of $i$ with probability
$1-p$.

\fig{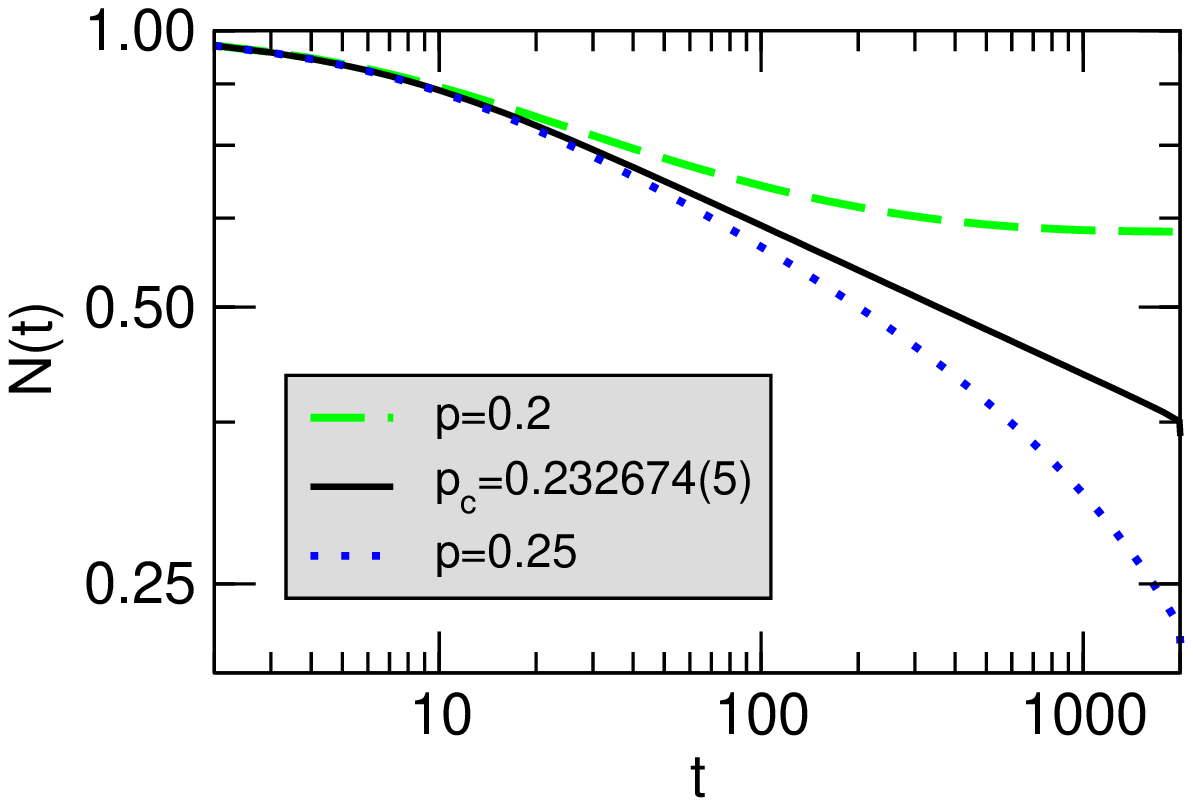}{fig:nt}{Density $N(t)$ of the $1D$ contact process 
in the absorbing phase ($p=0.25$), at criticality ($p=p_c$) and in the
ordered phase ($p=0.2$). Time is measured in units of the increment
$\Delta t=0.02$.}

There is a steady-state phase transition between the ordered phase
(`infected') at small $p$ (large $\lambda$) and the absorbing phase
(`healthy') at large $p$ (small $\lambda$).  To illustrate this,
consider the average density $N(t):=\langle n_i(t) \rangle$ which is
in fact independent of the lattice site $i$ and which is computed
using the methods described in section~\ref{sec:tmrg}. In
figure~\ref{fig:nt}, $N(t)$ is shown for three values of $p$.  There
is a critical value $p_c=0.232674(5)$ ($\lambda=\lambda_c=3.29785(8)$)
\cite{Hinr00} such that $N(t)\sim t^{-\delta}$. The exponent
$\delta\simeq 0.16$ can be read off from the slope in
figure~\ref{fig:mcrg}, in agreement with the literature \cite{Hinr00}.
For $p<p_c$ ($\lambda>\lambda_c$), the density converges exponentially
fast towards a positive steady-state value
$N_\infty:=\lim_{t\to\infty}N(t)>0$, while for $p>p_c$
($\lambda<\lambda_c$), the density decays exponentially fast towards
zero.

Correlations are defined as density-density autocorrelators, both in a
disconnected and a connected form
\BEA
C(t,s)      &:=& \left\langle n_i(t) n_i(s) \right\rangle \nonumber \\
\label{eq:gamma}
\Gamma(t,s) &:=& C(t,s) - N(t)N(s).
\EEA 
In order to obtain the autoresponse function
\begin{equation}
  \label{eq:resp}
  R(t,s)=\frac{\delta\langle n_i(t)\rangle}
  {\delta h_i(s)}\Bigg|_{h_i=0}
\end{equation}
we must introduce an external field $h_i$ coupled to the density
operator $n_i$ at site $i$.  The most straightforward possibility is
to introduce a spontaneous creation of particles
$\emptyset\longrightarrow A$ on site $i$ with rate $h_i$.

Previous experience with $R(t,s)$ comes from magnets, where the
system's response to an external field is studied via the integrated
response since data for $R(t,s)$ are usually too noisy.  For the
integrated response, the most common protocols in magnets are
zero-field-cooled (ZFC) and thermoremanent magnetisation (TRM), but
also intermediate protocols have been proposed. However, the analysis
of the time-dependent scaling of integrated response functions is not
always straightforward and the scaling behaviour of interest may even
be obscured by non-scaling terms or finite-time corrections. This
occurs in particular for quenches into the ordered phase
\cite{Henk03e}.  Another difficulty is that the application of a
uniform external field $h$ over several time steps introduces extra
particles into the lattice which brings the steady-state into the
active phase.  For magnets, sophisticated techniques have been
proposed \cite{Barr98,Plei03,Chat03,Ricc03} in order to avoid such a
systematic change of the system's properties and which may involve
random external fields which change sign in every time step, followed
by an average over realisations of the randomness or delicate changes
in the dynamics. Due to the absence of detailed balance in the
contact process, these methods are not available here.

In view of these difficulties, it is a great advantage of the LCTMRG
method that the autoresponse can be computed directly 
(see section~\ref{sec:tmrg}).
This is faster, more accurate and conceptually simpler.  The limit
$h\to 0$ can be taken analytically so the problems created by a
non-vanishing external field are circumvented.

In calculating $C(t,s)$, $\Gamma(t,s)$, and $R(t,s)$, we shall always
use a completely filled lattice as initial state.  This state is
completely uncorrelated and can be thought of as prepared at $p=0$
($\lambda=\infty$). 

Since the CP does not contain {\it a priori} a temperature variable,
we might try and define an analogy of the fluctuation-dissipation
ratio as follows 
\BEQ \label{eq:fdr}
  X(t,s) := \frac{R(t,s)}{\partial\Gamma(t,s)/\partial s}.
\EEQ

\section{The LCTMRG method}
\label{sec:tmrg}

Instead of the usual Monte Carlo (MC) simulation of the time-evolution
of the contact process, see \cite{Rama04}, 
we use a new variant of the DMRG applied to
stochastic transfer matrices, the so-called light-cone
transfer-matrix density-matrix renormalisation group 
(LCTMRG) \cite{Kemp03}.  This algorithm, which is an improvement on 
earlier DMRG approaches to calculate the time evolution of stochastic 
systems \cite{Kemp01,Enss01} has the
following advantages:
\begin{itemize}
\item There is no need for random numbers and ensemble averages since
  \emph{all} relevant ensembles and correlations (in the sense
  explained below) are taken into account.  One LCTMRG run takes a few
  minutes while 1000 MC runs may take days.  The resulting correlation
  functions are very smooth and require no further statistics, e.g.\
  in order to compute numerical derivatives with respect to time like
  $\partial \Gamma(t,s)/\partial s$.
\item The transfer matrix enables us to take the thermodynamic limit
  $L\to\infty$ \emph{exactly}.
\end{itemize}
However, the LCTMRG is still plagued by numerical instabilities whose
exact origin is unclear, restricting the calculation to about 1000
time steps.  The LCTMRG is not very useful for models where each site
may have many different states ($n\gg 2$), or where the interaction
spans more than two or three sites.

The dynamics of the one-dimensional stochastic process can be mapped
by a Trotter-Suzuki checkerboard decomposition onto a two-dimensional
classical model: this is the geometric interpretation of percolation
in two spatial dimensions, directed along one of the two axes.  
\begin{figure}[ht]
  \begin{center}
    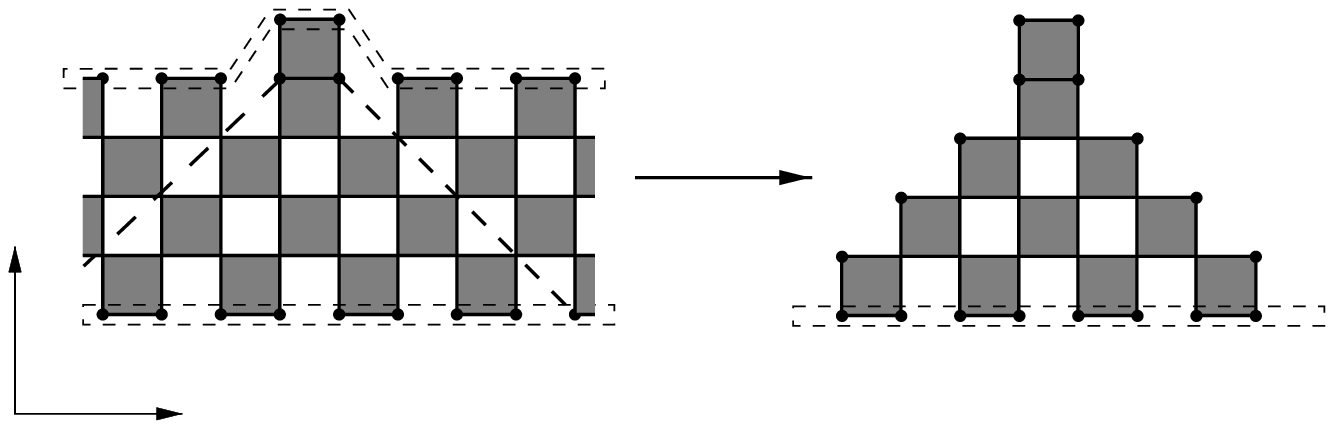
    \caption{(a) Trotter-Suzuki decomposition of $2\Delta t$ time
      steps. The resulting 2D lattice consists of local plaquette
      interactions $\tau$ and is infinitely extended in space direction.
      The dimension of the time direction is finite and the boundary
      conditions are fixed by $\bras{1}$ and $\cats{P(0)}$. (b)
      Reduction of the 2D lattice to a triangle structure. All other
      plaquettes trivialise, i.e.\ do not contribute to the state of 
      the top of the triangle. After \cite{Kemp03}.}
    \label{fig:trotter}
  \end{center}
\end{figure}
The checkerboard is made up of plaquettes $\tau$ (``local
transfer-matrices'') encoding the local interaction according to the
rules in figure~\ref{fig:cpupdate}:
\begin{equation}
  (\tau)_{r_1r_2}^{l_1l_2} = \bras{l_2r_2}e^{-\Delta t\cdot
    h}\cats{l_1r_1} = 
  \begin{minipage}{1.5cm}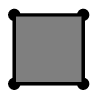\end{minipage}
  \quad \mbox{\rm with} \quad l_i,r_i\in\{0,1\}
\end{equation}
where $h$ is the local transition-rate matrix from two neighbouring
sites $l_1r_1$ at time $t$ to the same sites at time $t+\Delta t$.
The time step $\Delta t\ll 1$ should be chosen sufficiently small.

We determine the thermodynamic properties of the system by a transfer
matrix: this ensures that the system is truly infinite in space, while
we can follow the short-time dynamics for a certain number of time
steps.

Because of probability conservation (Eq.\ (\ref{eq:triv})) and
causality (at each time step, only a neighbouring site may be affected
by the local interaction), the measurement of a local observable
$n_i(t)$ at time step $t$ and site $i$ depends only on the ``past
light-cone'' of this site on the classical 2D lattice (\cite{Enss01};
see figure~\ref{fig:trotter}).
\begin{equation} \label{eq:triv}
  \forall l_1,r_1: \;\sum_{l_2 r_2} (\tau)_{r_1 r_2}^{l_1l_2}=1, \quad
  \begin{minipage}{2.6cm}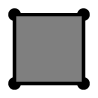\end{minipage}\hskip-8mm
  =1.
\end{equation}

\begin{figure}[htbp]
  \centering
  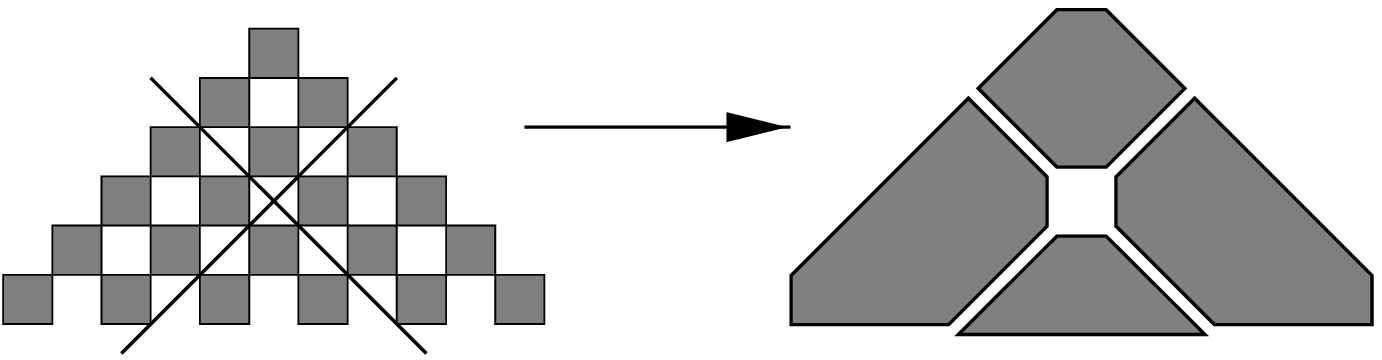
  \caption{Splitting the light-cone into four corner transfer-matrices
    (after \cite{Kemp03}).}
  \label{fig:ctm}
\end{figure}

As the dimension of the exact transfer-matrix grows exponentially with
the number of time steps, we use the density-matrix renormalisation
group (DMRG) idea to decimate the state space.  The DMRG relies on
splitting the system into two strongly correlated parts, called the
``system'' and ``environment''.  Kemper et {\it al.}\ \cite{Kemp03}
have proposed an efficient realisation of the DMRG algorithm applied
to corner transfer-matrices.  These are obtained by diagonal cuts
through the checkerboard: the light cone is split into four parts
diagonally along the future and past light cone of the center point of
the triangle (see figure~\ref{fig:ctm}).
For details we refer to \cite{Kemp03}.

We made a modification to the algorithm in order to be able to compute
the autocorrelation and autoresponse functions necessary for
investigating the ageing behaviour.  Usually for a local observable
$n_i(t)$ the expectation value is obtained by multiplying the local
transfer-matrix at site $i$ and time $t$ with $n_i(t)$ before applying
the initial and final conditions and taking the trace over temporal
indices.  For the two-time correlation function $C(t,s)$, the
algorithm has been modified to multiply the local transfer matrices
$\tau$ adjacent to site $i$ with $n_i$ both at time steps $s$ and $t$
before the trace.  From $C(t,s)$ the connected autocorrelation
$\Gamma(t,s)$ is computed via (\ref{eq:gamma}), and the derivative of
the connected autocorrelation function is computed from a symmetric
difference, i.e.\ 
\BEQ
  \label{eq:diffquot}
  \frac{\partial \Gamma(t,s)}{\partial s} :=
  \frac{\Gamma(t,s+\Delta t/2)-\Gamma(t,s-\Delta t/2)}{\Delta t}
\EEQ
which is sufficiently accurate (i.e.\ independent of $\Delta t$ for
$\Delta t=0.01\ldots 0.05$).

Likewise, when applying an external field $h_i$ in order to compute
$R(t,s)$, the local $\tau$ adjacent to site $i$ at time step $t=s$ is
modified to include particle production at rate $h_i$.  However, as we
are interested in the derivative with respect to the external field,
it is better to compute this derivative analytically: the Hamiltonian
in the presence of an external field $h_i$ on site $i$ is
\begin{eqnarray*}
  H_{h_i} = H + \id_{\dotsc,i-1} \otimes h_i 
  \begin{smallmatrix}
    -1&0\\
     1&0
  \end{smallmatrix}_i
  \otimes \id_{i+1,\dotsc}
\end{eqnarray*}
where $H$ is the stochastic Hamiltonian for the CP, see e.g.\
\cite{Schu00,Henk03r} for reviews.  Then using the
state at time $t=s$, $\ket{P(s)} = e^{-Hs}\ket{P(t=0)}$, and the
final state $\bra{1}$,
\newpage\typeout{ *** hier ist ein Seitenvorschub ! *** }
\begin{eqnarray*}
  R(t,s) & = \lim_{h_i,\Delta t' \to 0} \bra{1} e^{-H(t-s-\Delta t')}
  \left( \frac{e^{-H_{h_i}\Delta t'} - e^{-H\Delta t'}}{h_i \Delta t'}
  \right) \ket{P(s)} \\
  & = \lim_{h_i,\Delta t' \to 0} \bra{1} e^{-H(t-s-\Delta t')}
  \left( \frac{(H-H_{h_i})\Delta t' + \mathcal{O}((\Delta t')^2)}
    {h_i \Delta t'} \right) \ket{P(s)} \\
  & = \bra{1} e^{-H(t-s)}
  \left( \id_{\dotsc,i-1} \otimes
    -\begin{smallmatrix}
      -1&0\\
       1&0
     \end{smallmatrix}_i
    \otimes \id_{i+1,\dotsc} \right) \ket{P(s)}
\end{eqnarray*}
where the matrix is written in terms the local basis $(0,A)$ on site
$i$.  This has two advantages:
\begin{enumerate}
\item The limit $h_i\to 0$ is taken exactly, thus there is no danger
  of triggering a phase transition by inserting extra particles into
  the system.
\item No numerical derivative is necessary which would have included
  the difference of two very similar quantities, so this method is
  numerically more accurate.
\end{enumerate}

\fig{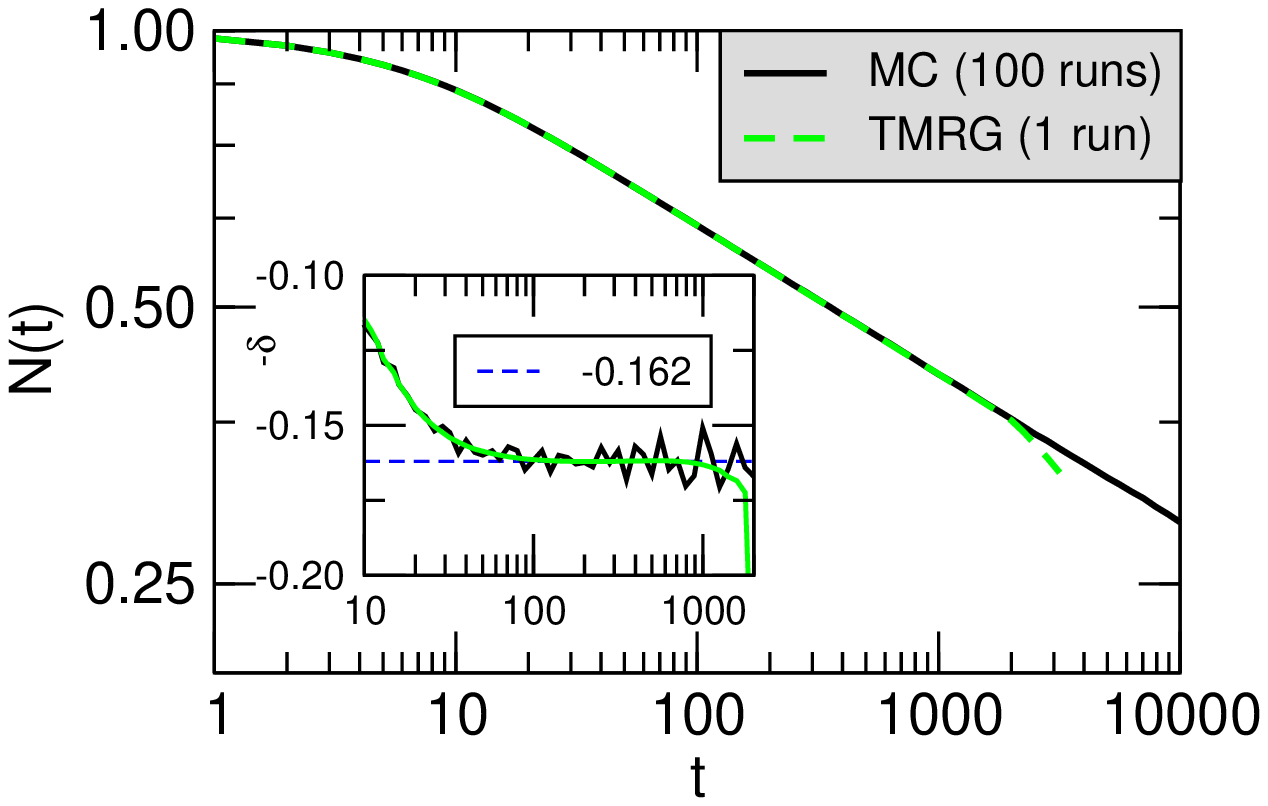}{fig:mcrg}{Comparison of the mean particle-density $N(t)$ found
by Monte Carlo (MC) and by the LCTMRG for the 
critical $1D$ contact process, $p=p_c$. The slope $-\delta$ is shown the
inset. The LCTMRG becomes numerically unstable around $t=1000$. Times are
measured in units of $\Delta t=0.02$.}

In figure~\ref{fig:mcrg}, we compare the results for $N(t)$ of the critical
contact process in $1D$ obtained from Monte Carlo (MC) or the LCTMRG, 
respectively. First, we observe that the LCTMRG data are fairly smooth, as is
especially evident when studying the exponent $\delta$ directly. However, 
we also see that the LCTMRG 
becomes numerically unstable around $t=1000$ time steps.  This happens
because the basis vectors of the reduced state space 
offered during the renormalisation by the LCTMRG method step become
inadequate: the expectation value of the identity operator $\langle 1
\rangle$ is around 1 (as it should be) only for the first several
hundred time steps, then decreases to below 0.1. However, the onset 
of instability can in practise always be identified very reliably. 
The reason for this instability is that DMRG works best if system and
environment are quite strongly entangled, which is not the case here. 

\section{Results}
\label{sec:results}

In this section, we first present the results of our numerical
calculations.  In the subsections \ref{sec:results:efftemp} and
\ref{sec:results:lsi}, we shall discuss two important implications of
our numerical results.
 
The MC results in figure~\ref{fig:mcrg} are obtained using the 3-site
update in the first row
of figure~\ref{fig:cpupdate} with asynchronous dynamics and parallel
update, system size $L=10^5$, $10^4$ time steps, and ensemble
averaging over 100 runs.  All LCTMRG results are obtained with the
2-site update shown in the second row of figure~\ref{fig:cpupdate},
using the following parameters: time step size $\Delta t=0.02$, number
of states in projected state space $m=32$, external field strength
exactly $h\to 0$, lattice size $L=\infty$ (exact thermodynamic limit).

\subsection{Ageing at criticality}
\label{sec:results:crit}

For the critical contact process, we now try and see whether the scaling
forms used to describe ageing in magnets apply. We consider the scaling forms
\begin{subeqnarray}
C(t,s) &=& s^{-b} f_C(t/s) \;\; , \;\; f_C(y)\sim y^{-\lambda_C/z} \label{Cs}\\
\Gamma(t,s) &=& s^{-b} f_{\Gamma}(t/s) \;\; , \;\; 
f_{\Gamma}(y)\sim y^{-\lambda_{\Gamma}/z} \label{Gs} \\
R(t,s) &=& s^{-1-a} f_R(t/s) \;\; , \;\; f_R(y)\sim y^{-\lambda_R/z} \label{Rs}
\end{subeqnarray}
where the asymptotic behaviour holds for $y\to\infty$. 
\fig{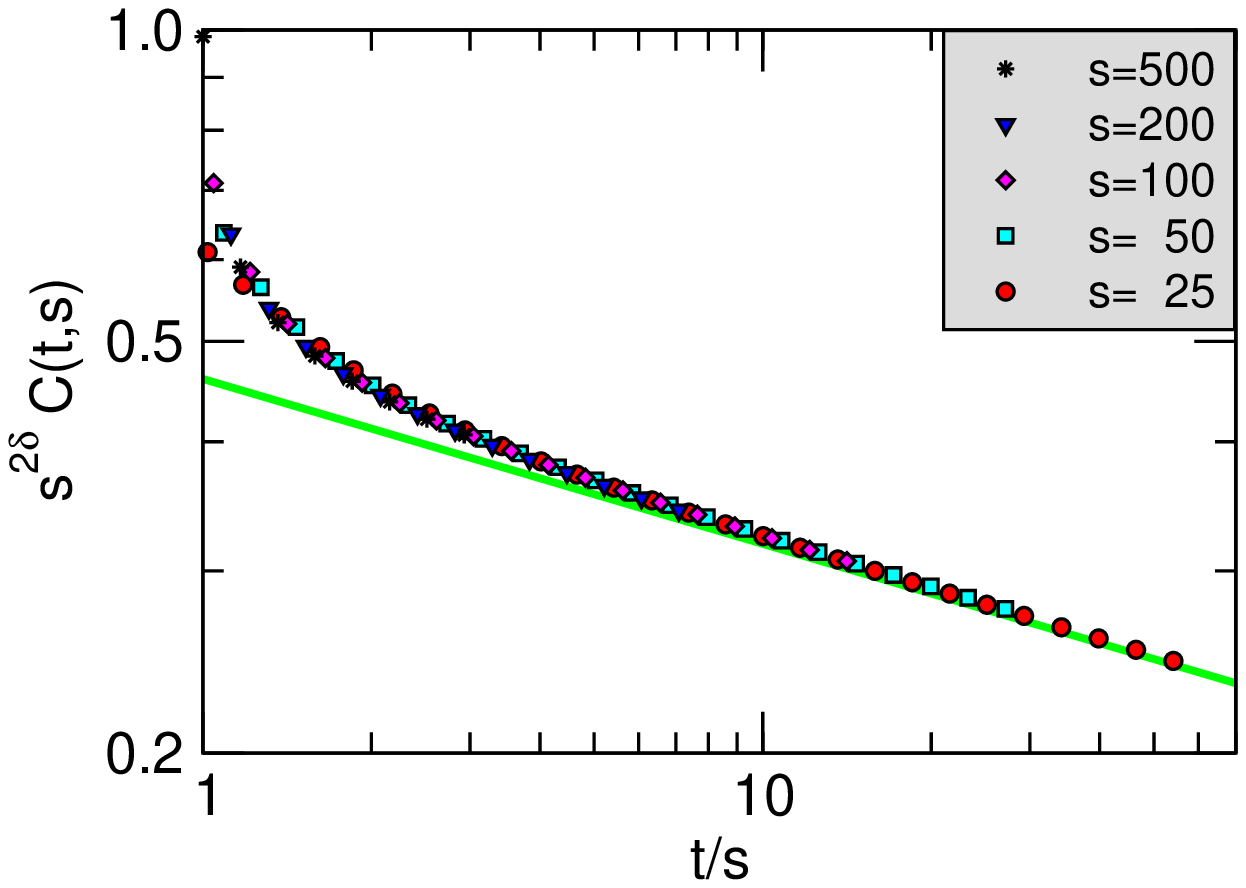}{fig:critC}{Autocorrelation $C(t,s)$ for the critical CP for
  several values of the waiting time $s$ as a function of the scaling variable
  $y=t/s$. The straight line has the slope $-0.16\approx -\delta$.}
First, we display in figure~\ref{fig:critC} the critical autocorrelator. A nice
data collapse is observed with an exponent $b=2\delta$ and for large values
of $t/s$, we find a power law according to eq.~(\ref{Cs}) 
with $\lambda_C/z=\delta$.\footnote{Here and in what follows, the numbers in
brackets give the estimated error in the last given digit(s).} 
This confirms the expectations derived in \cite{Rama04}. 

\fig{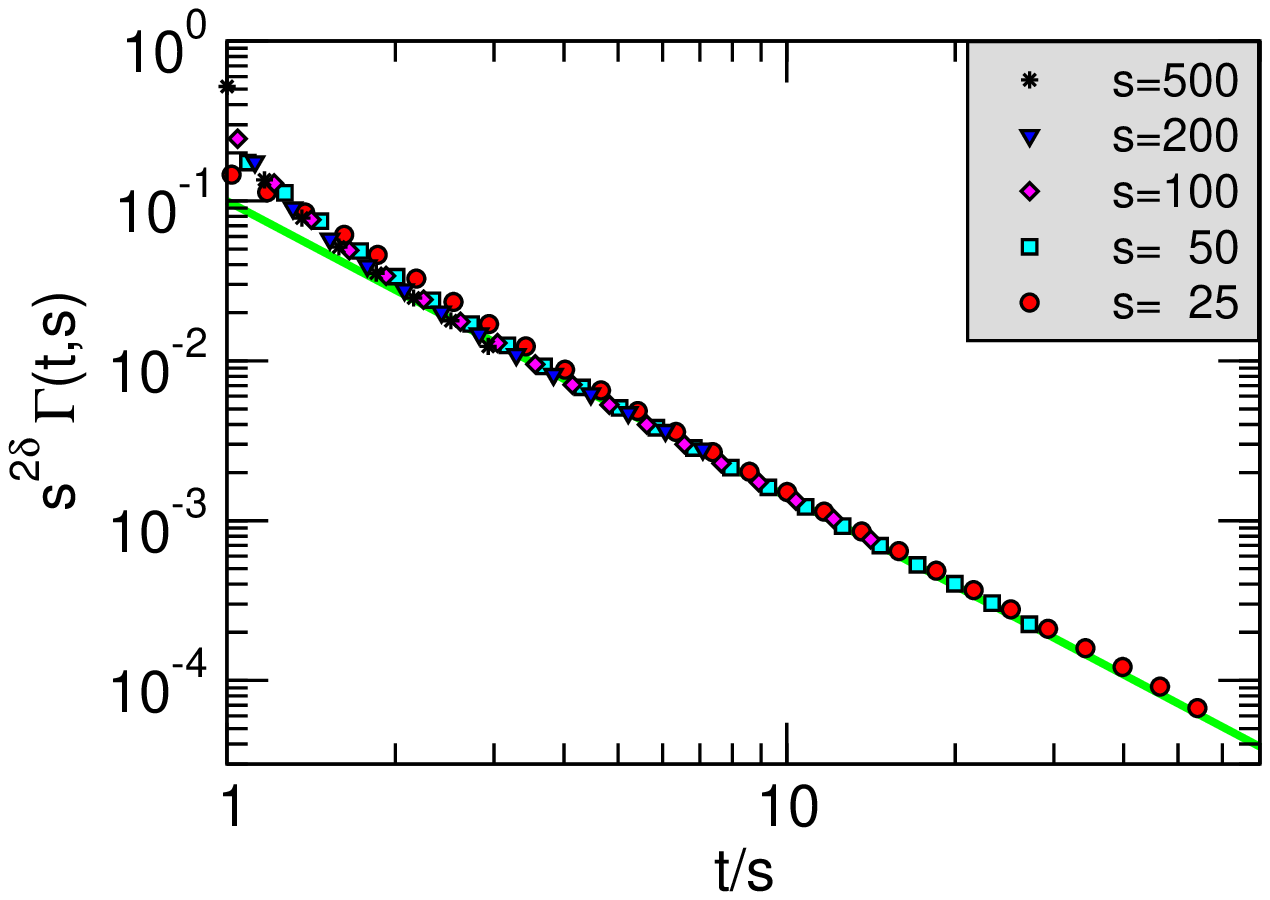}{fig:critG}{Connected autocorrelation $\Gamma(t,s)$ for the
critical CP, for several values of the waiting time $s$. The slope of the 
straight line is $-1.85$.}

Of course, we are interested in the deviations from this as measured by the
connected autocorrelator $\Gamma(t,s)$ which is shown in figure~\ref{fig:critG}. 
Again, the data collapse very well and we obtain the ageing exponents
$b=2\delta$ and $\lambda_{\Gamma}/z=1.85(10)$. 

\fig{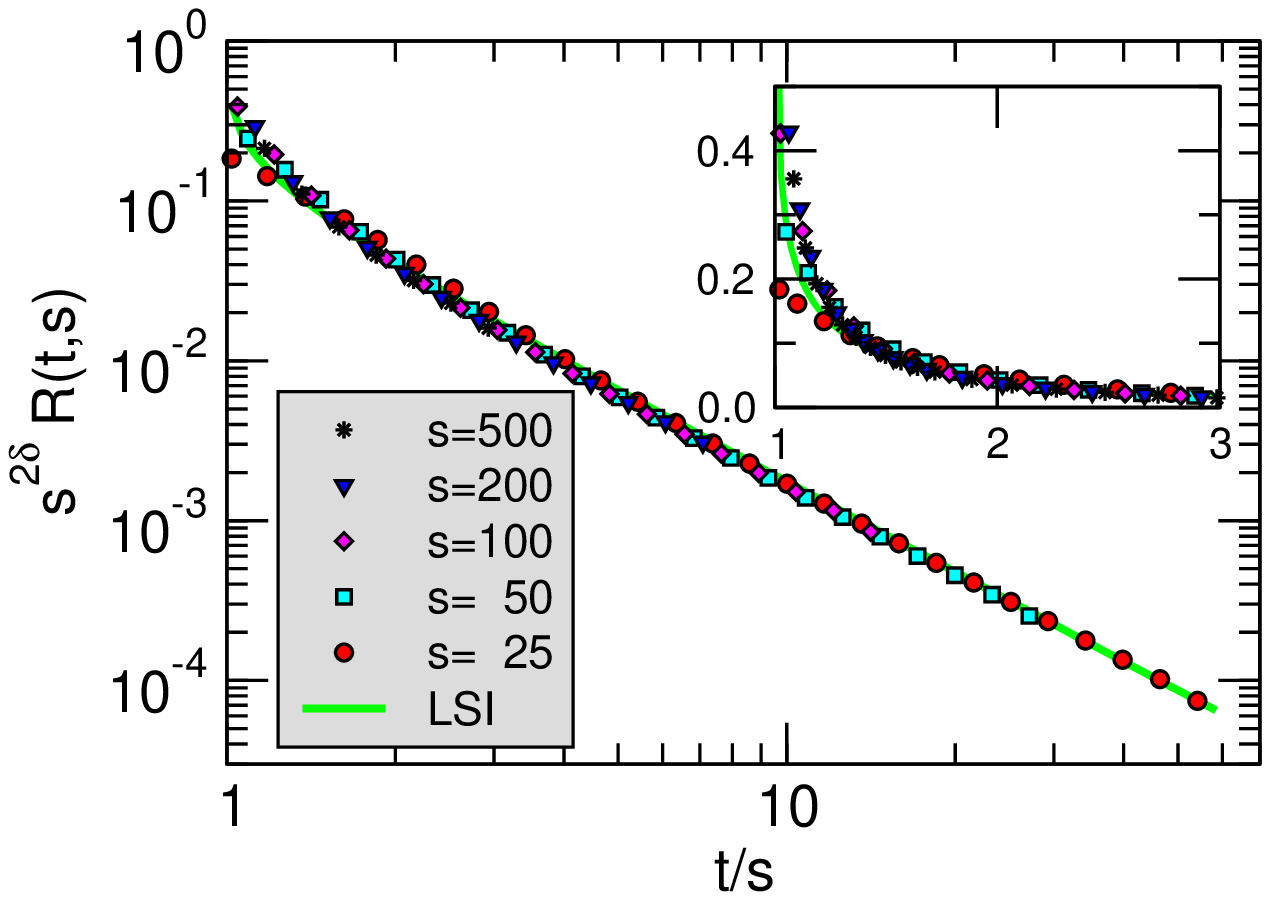}{fig:critR}{Autoresponse function $R(t,s)$ for the 
critical CP for several values of the waiting time $s$. The full curve labelled 
LSI is the prediction of local scale-invariance 
$f_R(y)= 0.12 \cdot y^{-1.85} (1-1/y)^{-2\delta}$. 
The inset shows the same data for $1\leq t/s\leq 3$.}

Next, we study the autoresponse function which is displayed in
figure~\ref{fig:critR}. As for the correlations, we find a neat
dynamical scaling behaviour and for large values of $y=t/s$, we can
read off the exponents $1+a=2\delta$ and $\lambda_R/z=1.85(10)$. We
shall come back in subsection~\ref{sec:results:lsi} to a quantitative
comparison of the precise functional form of the scaling function
$f_R(y)$ with the prediction of local scale-invariance.

\fig{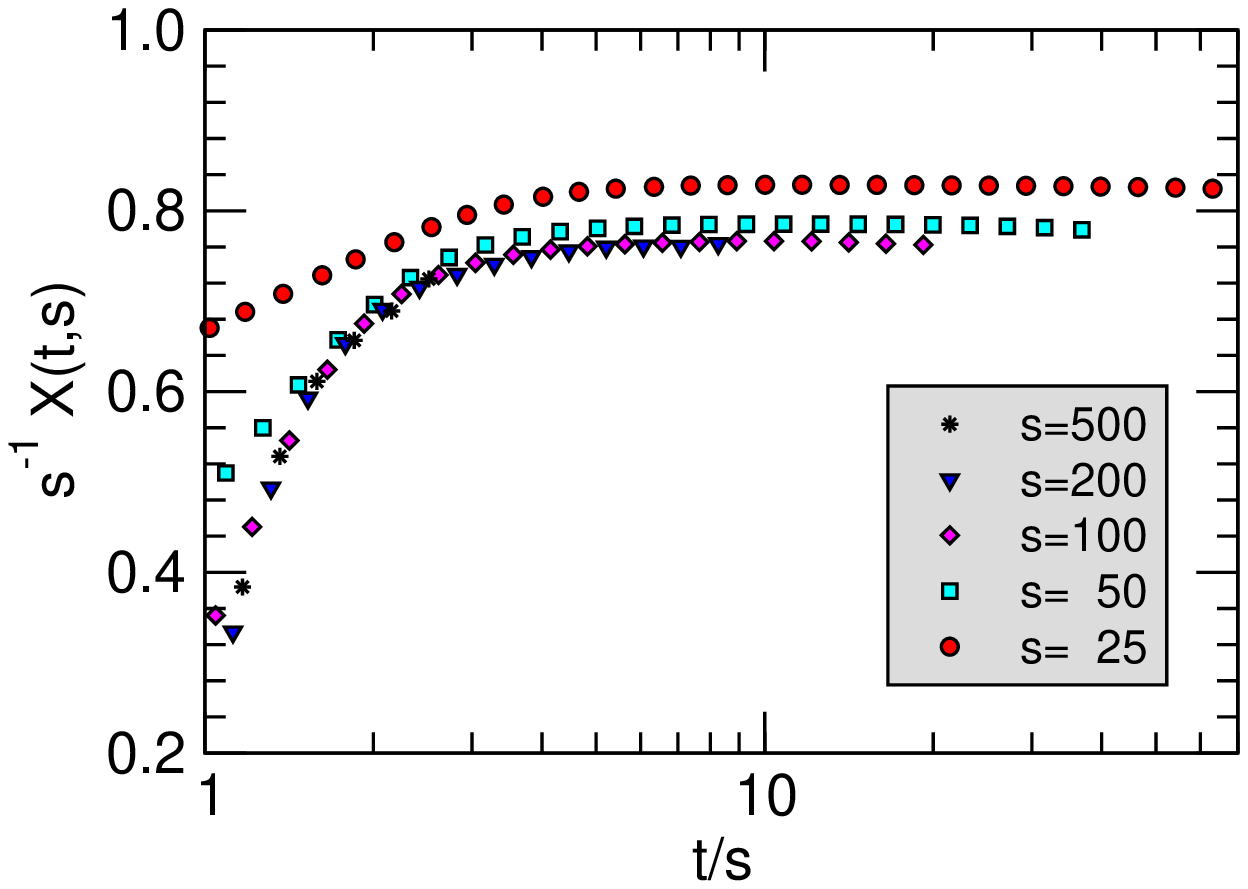}{fig:critX}{Fluctuation-dissipation ratio (FDR)
  $X(t,s)$ as defined in eq.~\protect{(\ref{eq:fdr})} for the critical
  contact process. For large waiting times $s\gtrsim 100$ the curves 
  $s^{-1}X(t,s)$ collapse and quickly saturate at $\approx 0.77(1)$.}

These tests provide evidence in favour of a dynamical scaling of the ageing
behaviour in the contact process which at least formally is quite analogous to
the one found previously in glasses and ferromagnets. However, there are 
qualitative differences between these models and the contact process as 
becomes explicit when considering the exponent relations
\BEQ
\lambda_{\Gamma} = \lambda_R \;\; , \;\;
1+a = b = 2\delta
\EEQ
In magnetic systems, the first of these had been simply taken for granted and
only recently, it has been established (i) that it need not hold for spatially
long-ranged initial correlations \cite{Pico02} or in certain glassy systems 
\cite{Sche03} and (ii) that for phase-ordering (hence $z=2$) with short-ranged
initial correlations local scale-invariance is sufficient to actually prove
the exponent equality $\lambda_{\Gamma}=\lambda_R$ \cite{Pico04}. On the other
hand, the {\em non-equality} of the ageing exponents $a$ and $b$ at a critical
point comes as a surprise. 
This important result can be illustrated in a different way by 
considering the analogue eq.~(\ref{eq:fdr}) of the fluctuation-dissipation
ratio. For the critical CP, we find from figure~\ref{fig:critX} for
$s\to\infty$ a scaling behaviour $X(t,s)\simeq s\cdot f_X(t/s)$ where the
scaling function $f_X(y)$ has a finite limit value $f_X(\infty)\simeq
0.77(1)$.

\subsection{Absorbing phase}
\label{sec:results:absorb}

Here, we discuss the behaviour in the subcritical or absorbing phase,
with $p=0.6>p_c$ ($\lambda<\lambda_c$). The asymptotic behaviour in this
phase may be understood by considering the extreme case $p=1$ first. For $p=1$,
the particles on different sites are uncorrelated and simply decay with a 
fixed rate. For any fixed site $i$ and with $t>s$, a moment's thought shows
that $n_i(t) n_i(s) = n_i(t)$ because $n_i\in\{0,1\}$. 
Therefore, $C(t,s)=N(t)$.

\fig{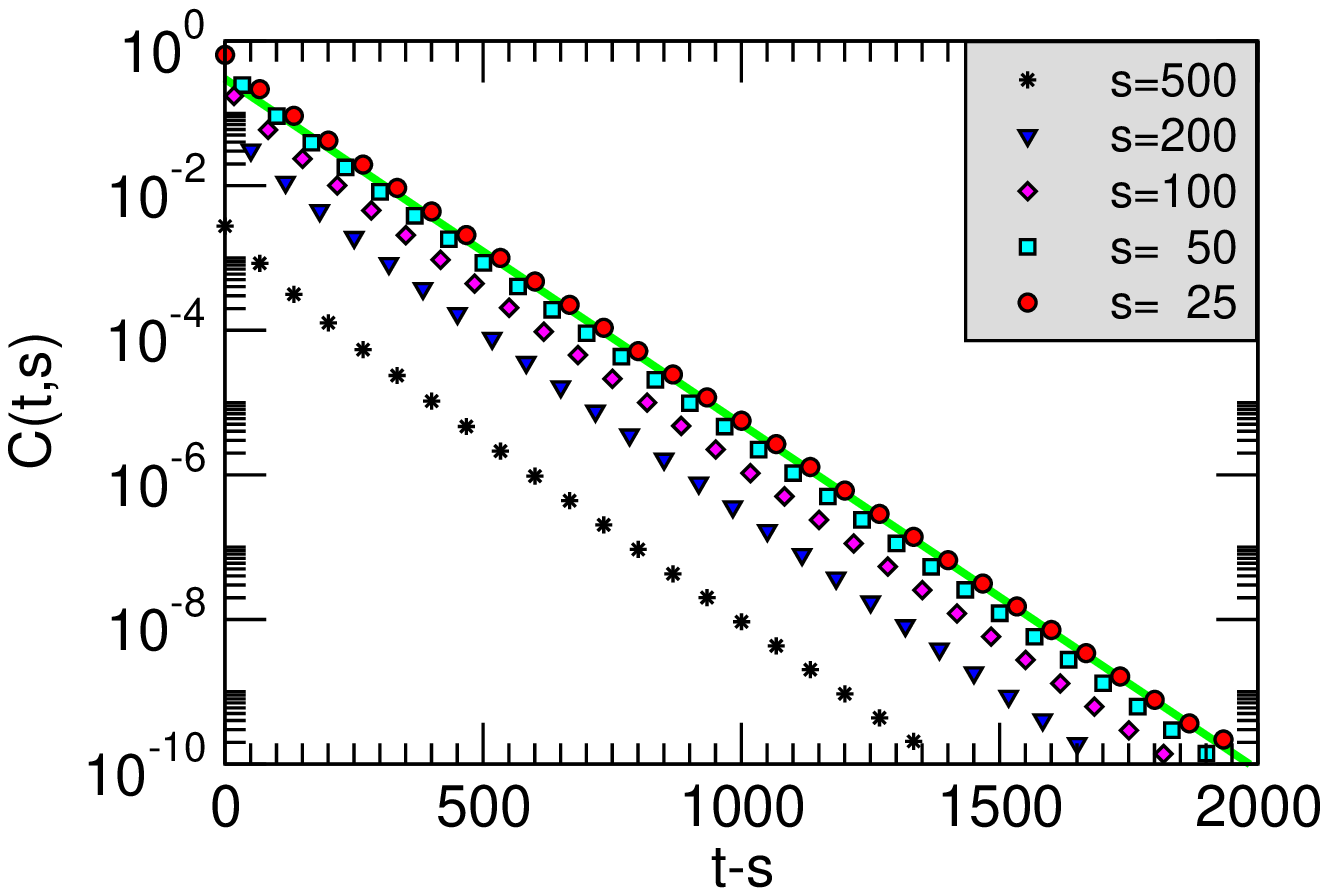}{fig:absorbC}{Autocorrelation $C(t,s)$ as a function of
  $\tau=t-s$ for several values of $s$ in the absorbing phase at $p=0.6$. The 
  full curve is proportional to $\exp(-0.01 (t-s))$.}

\fig{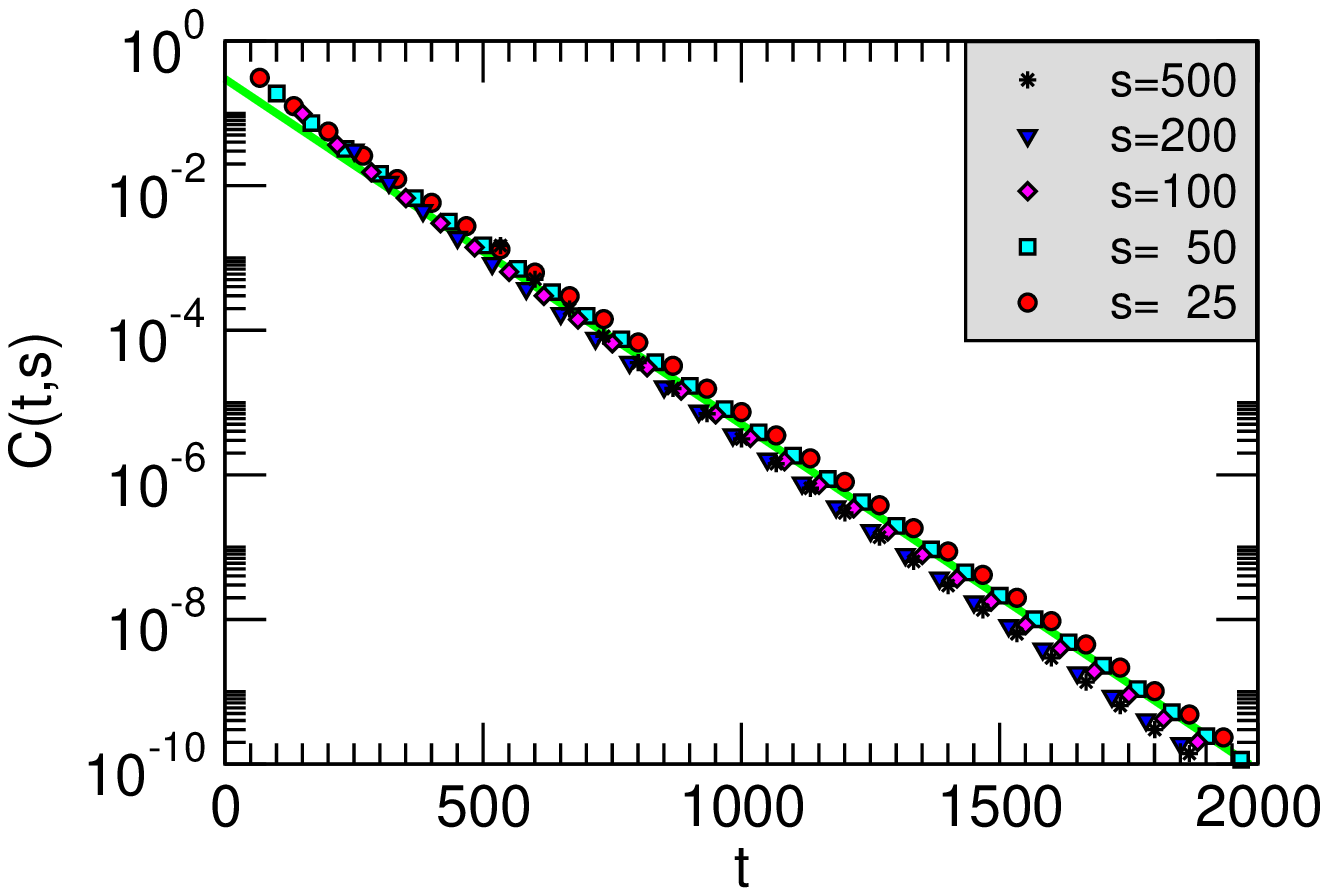}{fig:absorbCt}{Autocorrelation $C(t,s)$ as a 
  function of $t$ for the CP with $p=0.6$. The full curve is proportional 
  to $\exp(-0.01 t)$.}

This long-time behaviour of the autocorrelation survives in the entire 
absorbing phase, as is illustrated in figures~\ref{fig:absorbC} 
and \ref{fig:absorbCt} for $p=0.6$. 
We see that $C(t,s)$ decays exponentially fast but there is no collapse 
as a function of $\tau=t-s$ (we point out that for intermediate values of
$s$, the data appear to effectively collapse. For $p$ close to $p_c$, the
cross-over to the truly asymptotic behaviour may set in very late.). 
On the other hand, when plotted as a function
of $t$, there is a collapse for large values of $s$, see
figure~\ref{fig:absorbCt}. Similarly, $\Gamma(t,s)=C(t,s)-N(t)N(s)\sim N(t)$
for $s$ sufficiently large. 

\fig{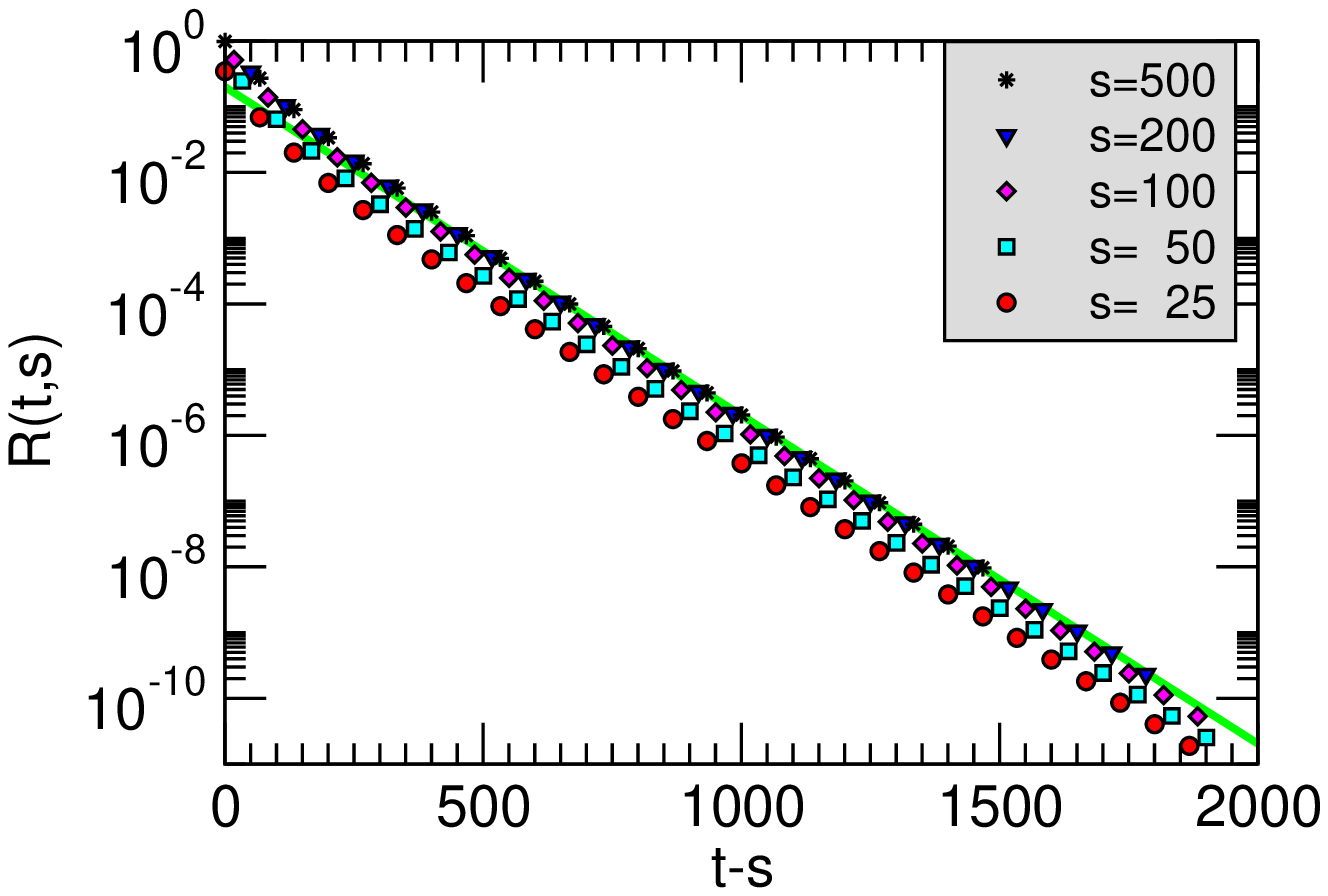}{fig:absorbR}{Autoresponse $R(t,s)$ as a function of
  $\tau=t-s$ for several values of $s$ in the absorbing phase at $p=0.6$. The 
  full curve is proportional to $\exp(-0.01 (t-s))$.}
On the other hand, for the response function we find in figure~\ref{fig:absorbR}
a rapid collapse in terms of the time difference $\tau=t-s$ for larger waiting 
times $s\gtrsim 100$ and time-translation invariance is recovered, as it 
should be. 

\subsection{Active phase}
\label{sec:results:active}

Finally, we discuss the long-time behaviour of the two-time
observables in the active (percolating/ordered) phase, where
$p=0.1<p_c$ ($\lambda=9>\lambda_c$). In distinction with magnetic
systems, where ageing also occurs for quenches into the ordered phase,
in the contact process time-translation invariance is rapidly
recovered and no ageing occurs.  We illustrate this in
figure~\ref{fig:activeG} for the connected autocorrelator and in
figure~\ref{fig:activeR} for the response function. In both cases,
when plotted against $\tau=t-s$, we observe a collapse for waiting
times $s\gtrsim 50$ and time-translation invariance is recovered.

\fig{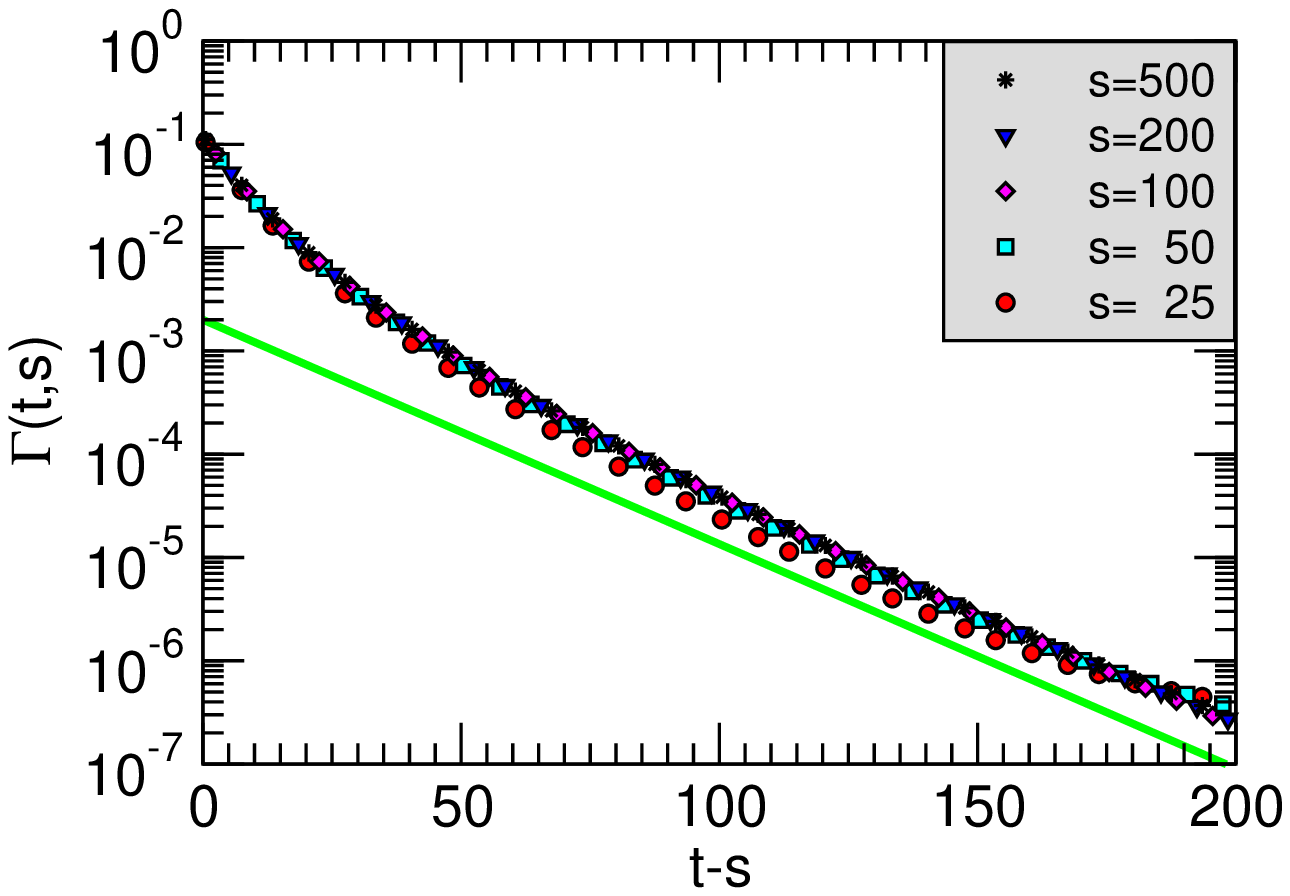}{fig:activeG}{Connected autocorrelation
  $\Gamma(t,s)$ in the active phase of the contact process (with
  $p=0.1$). The straight line is proportional to $\exp(-0.05 (t-s))$.}

\fig{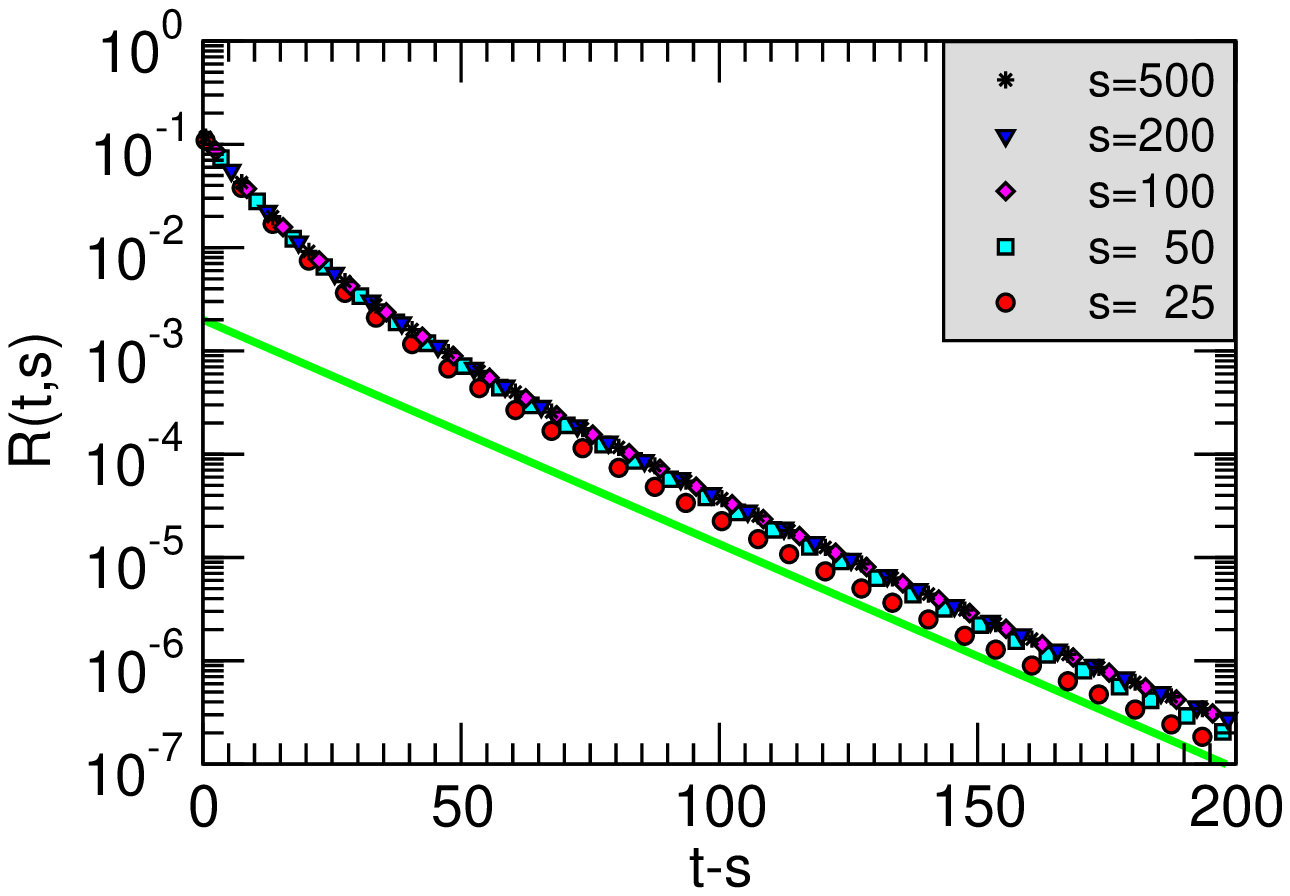}{fig:activeR}{Autoresponse $R(t,s)$ in the
  active phase of the contact process (with $p=0.1$). The straight
  line is proportional to $\exp(-0.05 (t-s))$.}

\subsection{Effective temperature}
\label{sec:results:efftemp}

One of the main peculiarities of equilibrium states with respect to more
general steady-states is that their probability distribution has the
form of a Boltzmann weight which may be 
characterised in terms of a temperature.
It therefore is a natural question whether for more general steady-states a
{\em non-equilibrium} temperature might be defined. Indeed, such an 
attempt has been presented recently by Sastre et {\it al.} \cite{Sast03}. 
They considered the fluctuation-dissipation ratio eq.~(\ref{1:gl:FDR}) and
observed that in the limit $s\to\infty$ and $t-s\to 0$, one should have
an equilibrium-like regime, hence $X(t,s)\to 1$. From this observation, they
define a \emph{dynamical temperature} by 
\BEQ \label{Tdyn}
\frac{1}{T_{\rm dyn}} := \lim_{t\to\infty}\left(\lim_{t-s\to 0}
\frac{R(t,s)}{\partial C(t,s)/\partial s}\right)
\EEQ 
By explicit calculation, they confirm that in the $2D$ critical voter model
this limit exists, has a non-trivial value and is universal
\cite{Sast03}. However, the basic assumption of the idea of Sastre et {\it al.} 
has been critically reexamined by Mayer and Sollich \cite{Soll04} who construct 
in the $1D$ Glauber-Ising model undergoing coarsening a defect-pair observable 
such that the fluctuation-dissipation ratio $X(t,s)\ne 1$ in the short 
time-regime (in particular they show $\lim_{s\to\infty} X(s,s)=3/4$). 

We now ask whether the definition (\ref{Tdyn}) \cite{Sast03} could be extended 
to the contact process. Obviously, for the contact process the 
connected correlator $\Gamma(t,s)$ must be used in eq.~(\ref{Tdyn}). 
Before we do so, it may be useful to recall some constraints on the ageing
behaviour which hold for relaxation towards equilibrium. Equilibrium states
are steady-states (hence time-translation invariant) which satisfy in addition
the fluctuation-dissipation theorem (thus $X(t,s)=1$). This last condition
may also be replaced by the Onsager symmetry condition
$\langle A(t) B(s)\rangle_{\rm eq} = \langle B(t)A(s) \rangle_{\rm eq}$
for any two observables $A$ and $B$;  which together with time-translation
invariance is enough to reproduce the 
fluctuation-dissipation theorem \cite{Cugl94b,Cris03}. 

We are interested in equilibrium critical dynamics. Combining the scaling
forms of section~\ref{sec:intro} and time-translation invariance, we expect
\BEQ
C(t,s) \sim \left( t-s\right)^{-b} \;\; , \;\;
R(t,s) \sim \left( t-s\right)^{-1-a}
\EEQ
Combining this with the FDT, it follows $a=b$. Therefore, the equality
$a=b$ is a necessary condition that the quasi-stationary state found when
$t-s\ll 1$ is indeed a quasi-equilibrium one. 

We can apply this argument to the quasi-steady state of the contact
process characterised by the connected autocorrelator $\Gamma(t,s)$
(instead of $C(t,s)$) and the autoresponse $R(t,s)$. In
subsection~\ref{sec:results:crit}, we have seen that dynamical scaling
eq.~(\ref{Gs}) holds with the ageing exponents satisfying
$1+a=b=2\delta$. Consequently, since a necessary condition for the
existence of a quasi-equilibrium regime is {\em not} satisfied in the
contact process, the definition (\ref{Tdyn}) of a non-equilibrium
temperature cannot be extended to this universality class. It is
remarkable to see that the presence or absence of detailed balance
leads to very different response behaviour, once the dynamical 
scaling-regime is reached. We conclude that the definition (\ref{Tdyn})
proposed in \cite{Sast03} is likely to reflect peculiar properties of
the critical voter model rather than being generic. 

\fig{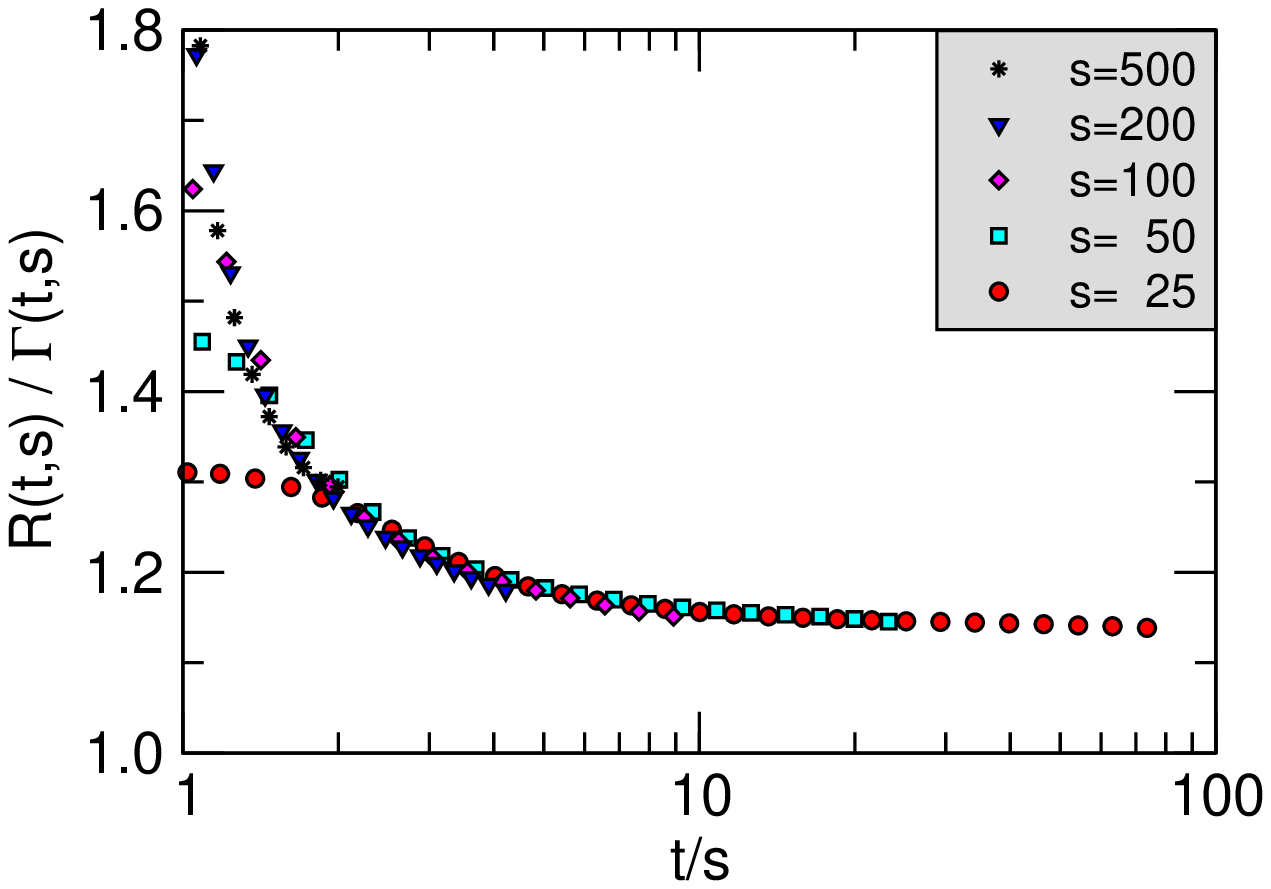}{fig:critRG}{Ratio $R(t,s)/\Gamma(t,s)$ for
  the critical CP.  For $s>25$ the curves appear to collapse to a scaling
  function which in turn saturates at a finite value for $t/s\gtrsim10$.}

Indeed, our data suggest that not the `fluctuation-dissipation ratio' $X(t,s)$
as defined in eq.~(\ref{eq:fdr}), but rather the ratio
\BEQ
\Xi(t,s) := \frac{R(t,s)}{\Gamma(t,s)}
\EEQ
should become a scale-invariant function of $t/s$ in the ageing regime at
criticality, viz.\ $\Xi=\Xi(t/s)$. This appears to be the case as is shown in 
figure~\ref{fig:critRG} and for $t/s\to\infty$ we read off the limit value 
$\Xi_{\infty}\simeq 1.15(5)$. In analogy of the universal limit 
fluctuation-dissipation ratio $X_{\infty}$ \cite{Godr00b} in systems with 
detailed balance, it would be interesting to see whether or not 
$\Xi_{\infty}$ is universal (for example by going beyond the mean-field
treatment \cite{Rama04} of Reggeon field-theory). 

\subsection{Local scale-invariance} 
\label{sec:results:lsi}

Having seen that at criticality, the ageing of the contact process satisfies
dynamical scaling, it is natural to inquire whether the recently proposed
extension \cite{Henk02} of dynamical scaling to a space-time-dependent, 
i.e.\ {\em local} kind of scale-invariance in magnets might also apply to the 
model at hand. The central assumption of that theory is that the linear
response function formed from so-called {\em quasi-primary} fields transforms
covariantly under the action of the group of local scale-transformations. 
We propose to try and see whether the form of the linear response in the 
critical contact process may be understood this way. 

Indeed, for any given value of $z$, infinitesimal local scale
transformations with a  space-time-dependent rescaling factor
$1+\eps(t,\vec{r})$ can be constructed.
In particular, the following explicit expression for the response function
is obtained from the condition that $R(t,s)$ transforms covariantly under
the action of local scale transformations \cite{Henk02,Henk01}
\BEQ \label{1:gl:lsi}
R(t,s) = r_0 \left(\frac{t}{s}\right)^{1+a-\lambda_R/z}
(t-s)^{-1-a}
\EEQ
and where $r_0$ is a normalisation constant. This prediction has been
confirmed in several models with a dynamics given by a master equation, 
notably the kinetic Ising model with
Glauber dynamics, both in the bulk \cite{Henk01,Henk03b} as well as close to
a free surface \cite{Plei04}, the kinetic XY model with a non-conserved
order parameter \cite{Abri04,Pico04} and for the Hilhorst-van Leeuwen model 
\cite{Plei04}. Local scale-invariance has also been confirmed for several 
variants of the exactly solvable spherical 
model and the free random walk \cite{Henk01,Godr00b,Cann01,Pico04}.

All these tests are based on a Master equation or on a linear Langevin
equation which reduces to a free field-theory. 
On the other hand, field-theoretical calculations of ageing ferromagnetic
systems based on non-linear Langevin equations find
small corrections to eq.~(\ref{1:gl:lsi}) \cite{Cala03,Maze04}. However, it
is not completely clear whether a description of ageing in terms of a Master
equation and in terms of a Langevin equation are completely equivalent.
A counterexample is provided by the $1D$ kinetic Ising model with 
Glauber dynamics at zero temperature where the universal exponent 
$\lambda_C=1$ is known exactly \cite{Godr00a,Lipp00}. 
On the other hand, the Langevin equation usually believed to be equivalent 
to this model, namely the time-dependent Ginzburg-Landau
equation, leads to the exact result $\lambda_C=0.6006\ldots$ \cite{Bray95}. 

Comparing eq.~(\ref{1:gl:lsi}) with the LCTMRG data for the $1D$ critical 
contact process, we find from figure~\ref{fig:critR} a perfect agreement 
almost down to $t/s=1$, where values for the exponents $a$ and $\lambda_R/z$ 
determined previously were used.\footnote{Recently, it has been established
that in those ageing systems which undergo cluster dilution rather than domain
growth, there is an universal early-time regime where dynamical scaling does
{\em not} hold \cite{Plei04b}. On the other hand, from Monte Carlo simulations
we know that in the contact process ageing proceeds via 
cluster dissolution \cite[Fig. 2]{Rama04}. Therefore a small deviation of the 
numerical data from the $t/s$ scaling for $t/s\approx 1$ is to be expected.}
The contact process hence provides the first example of a model satisfying 
local scale-invariance which has a non-equilibrium steady-state. 

\section{Conclusions}
\label{sec:conclusions}

We have studied the question if an analogue of the ageing phenomenon well-known
in magnetic systems also occurs in systems which relax to a non-equilibrium
steady-state. Our case study of the one-dimensional contact process has
allowed us to answer this general question affirmatively, at least in 
situations when the steady-state is critical and might be hence viewed as 
being formed from two coalescing steady-states. On the other hand, in the
active phase, where only a {\em single} stable steady-state exists, 
time-translation invariance is rapidly recovered and no ageing occurs. 
At criticality, the ageing behaviour can be described in terms of dynamical
scaling, see eq.~(\ref{Gs}), and we collect the values of the exponents in
table~\ref{tabelle1}. It is satisfying to see that in $1D$ we find a good 
agreement with the results of a Monte Carlo study \cite{Rama04} on the 
same model.\footnote{This agreement is not completely trivial, since 
the initial densities in these two studies are different.} While
we find evidence for the exponent equality $\lambda_{\Gamma}=\lambda_R$ in 
close analogy to magnetic systems quenched to criticality from a fully 
disordered state, the ageing exponents $a$ and $b$ are different and 
we conjecture 
\BEQ
1+a = b = 2\delta
\EEQ
where $\delta$ is a well-known non-equilibrium exponent. 
In particular, this implies that the quasistationary regime of the contact
process is already out of equilibrium and the recent attempt \cite{Sast03}
to define a genuine non-equilibrium temperature does not go through. Instead
of the fluctuation-dissipation theorem valid for equilibrium systems, we have
found evidence that in the contact process rather the ratio $R(t,s)/\Gamma(t,s)$
should converge to a finite value in the limit of widely separated
times (see figure~\ref{fig:critRG}). 

On the other hand, we have seen that the scaling form of the response function
is in agreement with local scale-invariance. We point out that this 
confirmation is obtained in a formulation based on the master equation and
not in a field-theoretical setting based on a Langevin equation. All existing
confirmations of local scale-invariance, see \cite{Henk04}, 
have either been obtained in this setup or else come from models which 
reduce to free field-theories.

\begin{table}
\begin{center}
\begin{tabular}{|c|llllll|c|} \hline \hline
$d$ & ~ & \multicolumn{1}{c}{$a$~} & \multicolumn{1}{c}{$b$~} 
    & $\lambda_C/z$ & $\lambda_{\Gamma}/z$ & $\lambda_R/z$ & method \\ \hline
1 & & \hspace{-3.3mm}$-0.68(5)$ & $0.32(5)$ & $0.16(1)$ & $1.85(10)$ & $1.85(10)$ &
    LCTMRG \\
  & & \hspace{-3.3mm}$-0.57(10)$ & $0.319$ & $0.159$ & $1.9(1)$ & $1.9(1)$ &
    Monte Carlo \\ \hline
2 &&  0.3(1) & $0.901(2)$ & 0.450 & $2.8(3)$ & $2.75(10)$ &
    Monte Carlo \\ \hline
$>4$ & & $\frac{d}{2}-1$ & 2 & 1 & -- & $\frac{d}{2}+2$ & 
    mean-field \\  
\hline\hline
\end{tabular}
\caption{Nonequilibrium exponents as defined in (\ref{Gs}) 
for the ageing of the critical contact
process in several space dimensions $d$, according to Monte Carlo \cite{Rama04}
or LCTMRG (this work) calculations. The exponents following from mean-field 
theory \cite{Rama04} are also included.\label{tabelle1}}
\end{center}
\end{table}

\ack

We thank A. Gambassi, M. Pleimling, J. Ramasco, M.A. Santos and 
C.A da Silva Santos for useful discussions and correspondence.


\end{document}